\documentclass[amsmath,amssymb,amsfonts,aps,prstper,citeautoscript,10pt,twocolumn]{revtex4-1}
\usepackage{graphicx}
\usepackage{multirow}
\usepackage{booktabs}
\usepackage{array}
\usepackage{amsmath}
\usepackage{url}

\begin{document}
\title{Benford's Law: Textbook exercises and multiple-choice testbanks}

\author{Aaron D. Slepkov}
\email{aaronslepkov@trentu.ca}
\affiliation{Trent University, Department of Physics and Astronomy, Peterborough, ON K9J 7B8, Canada}

\author{Kevin B. Ironside}
\affiliation{Trent University, Department of Physics and Astronomy, Peterborough, ON K9J 7B8, Canada}

\author{David DiBattista}
\affiliation{Department of Psychology, Brock University, St. Catharines, Ontario, L2S 3A1, Canada}

\begin{abstract}
%\vspace{1in}
Benford's Law describes the finding that the distribution of leading (or leftmost) digits of innumerable datasets follows a well-defined logarithmic trend, rather than an intuitive uniformity. In practice this means that the most common leading digit is 1, with an expected frequency of 30.1$\%$, and the least common is 9, with an expected frequency of 4.6$\%$. The history and development of Benford's Law is inexorably linked to physics, yet there has been a dearth of physics-related Benford datasets reported in the literature. Currently, the most common application of Benford's Law is in detecting number invention and tampering such as found in accounting-, tax-, and voter-fraud. We demonstrate that answers to end-of-chapter exercises in physics and chemistry textbooks conform to Benford's Law. Subsequently, we investigate whether this fact can be used to gain advantage over random guessing in multiple-choice tests, and find that while testbank answers in introductory physics closely conform to Benford's Law, the testbank is nonetheless secure against such a Benford's attack for banal reasons.  
\end{abstract}

\maketitle
\section{Introduction}\label{sec:intro}
The expectation that the leading digits in the numbers one encounters in everyday life are uniformly distributed is a well-known ``pitfall of elementary statistics" \cite{goudsmit_1944}. As early as 1881, mathematician Simon Newcomb observed that the first pages in his bound logarithm tables were considerably more worn than the latter pages in the book \cite{newcomb_1881}. This led him to posit that the leading digits of various numbers he encountered were logarithmically-distributed. Fifty years later, physicist Frank Benford independently observed the same effect in his log tables and came to the same conclusion. Benford then went on to collect over 20,000 numbers from 20 different datasets such as front-page newspaper entries, street addresses of eminent scientists, physical constants, and American League baseball statistics, to show that this finding applies to a wide range of phenomena \cite{benford_1938}. Benford's proposed distribution of leading digit frequencies is given by
\begin{equation} \label{eq:benford}
\text{Prob}\left(D_i\right)=\log\left(\frac{D_i+1}{D_i}\right) ;\;\;\; D_i\in \left\{1,2,3,...,9\right\},
\end{equation}
where Prob($D_i$) is the probability of finding $D_i$ as the leading digit in a given number. Therefore, the frequency of leading digits diminishes monotonically from 30.1\% and 17.6\% for the digits 1 and 2, to 5.1\% and 4.6\% for the digits 8 and 9, respectively. Benford called this finding a ``law of anomalous numbers" \cite{benford_1938}. Nonetheless, (in accordance with Stigler's Law of eponymy \cite{stigler_1980}) this has come to be known as Benford's Law of leading digits. \\

Must there be a well-defined distribution of leading digits for a given phenomenon or dataset? Intuitively, most people suspect that the most likely distribution of leading digits should be a uniform distribution, such that they are all equally likely. However, for a given physical phenomenon, any such stable distribution should be scale-invariant in a way that the choice of unit system or choice of base would maintain the existence of such a distribution. For example, if there is to be a well-defined distribution of leading digits for a tabulation of the mass of various insect species, it shouldn't matter if the mass is measured in milligrams or in grains of rice. It has been shown that Benford's Law gives the only scale invariant distribution of leading digits \cite{pinkham_1961}. Furthermore, such scale invariance assures base invariance as well \cite{hill_1995}. Thus, it could be said that if a given dataset is to have an identifiable leading-digit distribution, then it must follow Benford's Law \cite{raimi_1969}. However, there is no requirement that all datasets must have a stable distribution of leading digits, and thus not all phenomena follow Benford's Law.

\par Decades of research have led to some guidelines for predicting which datasets might follow Benford's Law \cite{nigrini_2012, berger_2011}. Such phenomena should span several orders of magnitude, and preferably should be unbounded. The numbers in the dataset should be physically-relevant and should represent a measurement that has associated units; phone numbers, lottery numbers, and license plate numbers are not expected to be Benford-distributed. Finally, it has been suggested that the dataset should contain more small numbers than large numbers\cite{lemons_1986, nigrini_2012}. Despite such guidelines, there are numerous examples of sets of numbers that violate at least one of these guidelines and still follow Benford's law, such as the Fibonacci Numbers \cite{wlodarski_1970}, or most geometric series \cite{benford_1938}. There are now numerous published examples of mathematical series and physical datasets that follow Benford's Law \cite{benfordonline}. These include geological streamflow rates \cite{nigrini_2007}, radioactive half-lives \cite{buck_1993}, hadron widths \cite{shao_2009}, statistical physics distributions \cite{shao_2010}, auction prices \cite{giles_2007}, and business invoices and tax returns \cite{nigrini_2012, busta_1992}, as well as the original datasets by Benford \cite{benford_1938}. The precision with which various types of data follow Benford's Law has led to its widespread application for forensic accounting and auditing \cite{nigrini_2012, hickman_2010, durtschi_2004}. Benford's Law has not yet found robust application in the sciences, although from early on there have been suggestions that it could be used for diagnostic applications in computer design and scientific calculation errors \cite{pinkham_1961, hill_1995}. Indeed, there has often been a sense that Benford's Law is purely mathematical, rather than physical, and will never find practical applications \cite{raimi_1985}.

\par Because only certain datasets that correspond to real phenomena closely follow Benford's Law, we trust that knowledge of this fact could be applied towards pro-active or predictive ends---as opposed to \emph{a posteriori} diagnostics. In consideration of the general guidelines for datasets that are expected to conform to Benford's Law, we identify the set of answers to end-of-chapter questions in introductory physics textbooks as a possible Benford's Law candidate. In this article, we test this hypothesis, and find the set of introductory science textbook answers to conform well to the logarithmic distribution of leading digits.

\par Confirmation that quantitative answers to a wide range of ``problems" in physics and chemistry follow Benford's Law immediately suggests a practical (if somewhat nefarious) application: Examination questions that are meant to assess knowledge of introductory physics and chemistry should be similar to those found in the practice sections of the textbooks. Thus, multiple-choice testbanks that are created to provide a valid examination tool might also follow similar leading-digit trends found in the textbook exercises. If this is true, then perhaps the implication that over 50\% of the answers to numerical multiple-choice exam questions are anticipated to have a leading digit of 1, 2, or 3 can be utilized to gain an advantage by physics-ignorant but test-wise students? Traditionally, a physics-ignorant student can resort to complete guessing on a multiple-choice test, and thus each question has an expected baseline for guessing that is based on the number of options in the question. A student has a 33\% chance of getting any given question right in a 3-option question, a 25\% chance in a 4-option question, and a 20\% chance in a 5-option question. Students employ a slew of test-wise strategies to boost this baseline \cite{testwise_strategies}. One such strategy---which is often scoffed by professors but may have some marginal advantage \cite{attali_2003}---involves preferential selection of middle options. This manifestation of edge-avoidance is colloquially known as ``when in doubt, choose C!". This strategy (as well as most others) is based on poor test construction, and is easily foiled. On the other hand, a Benford's Law-based attack on a multiple-choice testbank would be guided by the assumption that it is more likely that the incorrect options (distractors) are semi-randomly selected in such a way that each yields a uniform first-digit distribution. Thus, after demonstrating that answers to end-of-chapter textbook questions yield a Benford distribution, we discuss considerations for which a Benford's Law-based attack on a multiple choice testbank is expected to gain an advantage over random guessing. We then proceed to analyze the distribution of leading digits in an actual introductory physics multiple-choice testbank, and go on to find that when we attack the testbank directly the ubiquity of the Benford distribution in fact secures the bank against such attacks. 

\par Finally, we discuss how the rounding off of numbers to a pedagogically-motivated reduced set of significant figures is expected to alter the distribution of leading digits away from Benford's Law and demonstrate that a modified distribution based on this fact yields even better fits to testbank data in physics.

\section{Methods and Results}
\subsection{Data Collection}
To select a representative sample of physics (and chemistry) exercises, three books were chosen based on ready availability on our bookshelf on current popularity in undergraduate physics education: ``Physics for Scientists and Engineers: A Strategic Approach", 3\textsuperscript{rd} edition, by Knight (Pearson Education, 2013) is a highly popular introductory physics textbook with an approach that is based on recent physics education research findings. ``Sears and Zemansky's University Physics", 10\textsuperscript{th} edition, by Young \& Freedman (Addison Wesley Longman, 2000) was a popular calculus-based introductory physics textbook a decade ago, with emphasis on physics education fundamentals. ``Fundamentals of Analytical Chemistry", 7\textsuperscript{th} edition, by Skoog, West, and Holler (Saunders College Publishing, 1996) was a favorite intermediate-level undergraduate chemistry textbook for one of us (ADS), and was selected in anticipation of useful quantitative chemistry end-of-chapter questions. Henceforth these three books will be referred to as Knight, Young \& Freedman, and Skoog, respectively. The data was obtained by recording the leading digit (i.e. the leftmost nonzero digit) from every end-of-chapter answer. Data collection was implemented manually by parsing through the texts, but with a protocol designed to eliminate subjective selection. Nonetheless, because of the general constraints of anticipated Benford data sets, such as avoiding unphysical numbers and numbers that are too narrowly confined in domain, we rejected all unitless values, values reported as percentages, and those with units of degrees. Furthermore, because the number zero is meaningless from a significant digit standpoint, all answers of exactly zero were rejected. Obviously, non-numeric end-of-chapter answers such as pictures, graphs, equations, and textual answers were ignored. In all, approximately 10\%-15\% of entries in the introductory physics texts were rejected and 30\%-35\% were ignored as non-numeric. In Skoog---the analytical chemistry text---as many as 25\% of the numeric answers were rejected, largely due to being unitless or percentages. Other than these limitations, all other data was recorded. For testbank data, we recorded leading digits of the multiple-choice numerical items supplied with Knight. When parsing this testbank we looked at both the keyed options and at the distractors, recording the leading digit of each separately. In recording testbank entries, the same protocol was followed as described for textbook data, however the number of rejected entries in the testbank was only 7\%. In sum, this data is presented in Table~\ref{tab:data}.

\begin{table*}
\caption{Obtained distributions of leading digits and measures of conformation to Benford's Law for three textbooks, a multiple-choice testbank, and aggregate data}
\bgroup
\def\arraystretch{1.1}
{\setlength{\tabcolsep}{0.5em}
\begin{tabular}{l l || p{1.5cm} | p{1.5cm}  p{1.4cm}  p{1.4cm} | p{1.7cm}  p{2.0cm} | p{1.5cm}  }
 \hline \hline
\textbf{} & \textbf{} & \textbf{Benford dist.} & \textbf{Knight 3rd ed. end-of-chapter answers} & \textbf{Young 10th ed. end-of-chapter answers}  & \textbf{Skoog 7th ed. end-of-chapter answers} & \textbf{Knight testbank answers} & \textbf{Knight testbank distractors} & \textbf{combined data} \\
 \hline \hline
\multicolumn{2}{c||}{\textbf{\# entries}} & {} & 1644 & 2155 & 294 & 1485 & 5671 & 11249\\
 \hline
 \multirow{9}{*}{Leading Digit} & 1 & 0.3010 & 0.289 & 0.286 & 0.313 & 0.277 & 0.291 & 0.288\\
 
{} & 2 & 0.1760 & 0.167 & 0.169 & 0.167 & 0.187 & 0.182 & 0.177\\
 
 {} & 3 & 0.1249 & 0.125 & 0.113 & 0.122 & 0.143 & 0.127 & 0.125\\
 
 {} & 4 & 0.0969 & 0.099 & 0.110 & 0.112 & 0.088 & 0.101 & 0.101\\

{} &  5 & 0.0791 & 0.080 & 0.089 & 0.071 & 0.071 & 0.089 & 0.085\\
 
{} & 6&  0.0669 & 0.080 & 0.073 & 0.065 & 0.084 & 0.062 & 0.070\\
 
{} & 7 & 0.0579 & 0.057 & 0.062 & 0.051 & 0.061 & 0.053 & 0.056 \\
 
{} &  8 & 0.0511 & 0.054 & 0.060 & 0.034 & 0.051 & 0.047 & 0.051 \\
 
{} & 9 & 0.0457 & 0.049 & 0.038 & 0.065 & 0.039 & 0.050 & 0.046 \\
% \hline
% $\mathbf{\chi ^2}$ \footnote{$\mathbf{\chi ^2}$ threshold for rejecting null hypothesis (p \textless 0.05) is 15.5} & {} & {} & 6.69 & 19.40 & 5.48 & 17.77 & 19.37 & 14.87 \\
 \hline
  \textbf{MAD}\footnote{MAD = Mean Absolute Deviation (see text for definition)} & {} & {} & 0.0050 & 0.0094 & 0.010 & 0.011 & 0.0054 & 0.0033 \\
 \hline 
  \textbf{Conform to BL?}\footnote{BL = Benford's Law} & {} & {} & close conform & accept. conform & accept. conform & accept. conform & close conform & close conform \\
 \hline \hline
 \end{tabular}}
\egroup
\label{tab:data}
\end{table*}

\begin{figure}
 \includegraphics[width=\columnwidth]{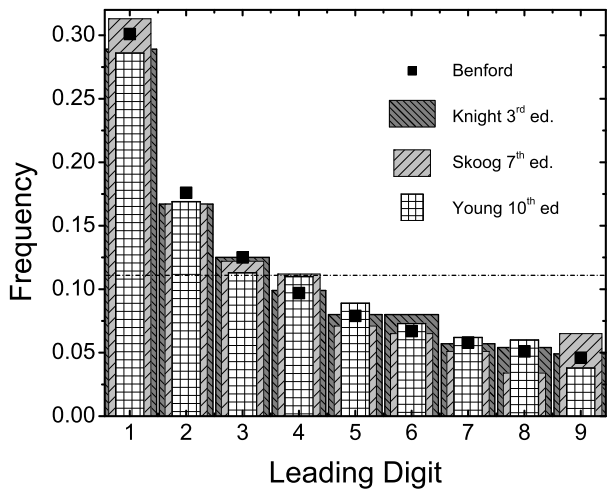}
 \caption{The distribution of leading digits in end-of-chapter excercise answers from two popular introductory physics textbooks (Knight, Young \& Freedman) and an analytical chemistry textbook (Skoog). The dashed horizontal line indicates uniform distribution of first digits. Statistical analysis confirms conformation to Benford's Law, overlayed as black squares.}\label{fig:textbooks}
\end{figure}

\subsection{Data Analysis}
The data presented in Table~\ref{tab:data} includes the theoretically-expected leading digit distributions of Benford's Law.  According to Eqn. \ref{eq:benford} this distribution can never be perfectly realized in any dataset because the values are irrational numbers. Thus, Benford's Law can only be \emph{approached}, and any dataset---no matter how good---will ultimately deviate from the ideal distribution. In many cases, where the invocation of Benford's Law is meant simply to highlight the disproportionate abundance of low-value leading digits over high-value leading digits, the suggestion of Benford's Law can be confirmed by visual inspection of a digit frequency histogram. Such a histogram is presented in Fig.~\ref{fig:textbooks}, showing the distributions for end-of-chapter exercise answers from the three textbooks. All three textbooks clearly yield a Benford-like distribution with an observed monotonic decrease in frequency with increasing leading digit value. Establishing a benchmark statistical measure of conformity to Benford's Law has been an ongoing research endeavor \cite{nigrini_2012, giles_2007, alexander_2009}. In the physics Benford's Law literature, the most common measure of statistical conformity to a Benford distribution is the $\chi^2$ test for goodness of fit. However, leading expertise in Benford's Law analysis finds that the $\chi^2$ test suffers from an ``excess power" problem, wherein larger data sets require increasingly better fits to pass the $\chi^2$ threshold for conformity \cite{nigrini_2012}. Thus, larger data sets that by inspection give better fits than smaller datasets will often fail a $\chi^2$ test that the smaller dataset passes \cite{chi2_note}. 

As a way to avoid the excess power problem of the $\chi^2$ test, Nigrini has proposed using a mean absolute deviation (MAD) measure as the benchmark test for dataset conformity to Benford's Law \cite{nigrini_2012}. MAD is an empirically-based whole-test measure that simply takes the average of the absolute deviation of each digit's frequency from the ideal Benford's Law frequency. Specifically, this is given by
\begin{equation} \label{eq:MAD}
\text{MAD}=\frac{\sum\limits_{i=1}^K |AP-EP|}{K},
\end{equation}
where $K$ is the number of leading digit bins (9 for first leading digit; 90 for first two leading digits, etc.), AP is the actual proportion observed, and EP is the expected proportion according to Benford's Law. The MAD test does not have an analytically-derived critical value. Instead, Nigrini has established empirically-based criteria for conformity to Benford's Law.\cite{nigrini_2012} The suggested MAD ranges for ``close conformity", ``acceptable conformity", and ``marginal conformity" are $0.000-0.006$, $0.006-0.012$, and $0.012-0.015$, respectively. A MAD value above $0.015$ is considered non-conforming.

\subsection{Results}
As seen from Table \ref{tab:data}, Knight closely conforms to Benford's law (MAD=0.0050), while Young \& Freedman (MAD=0.0094) and Skoog (MAD=0.010) both show acceptable conformity. A visual inspection of the distribution histogram presented in Fig. \ref{fig:textbooks} also strongly suggests close conformation. Thus we conclude that, in general, numeric end-of-chapter questions in physics and chemistry textbooks follow Benford's Law.

\begin{figure}
\includegraphics[width=\columnwidth]{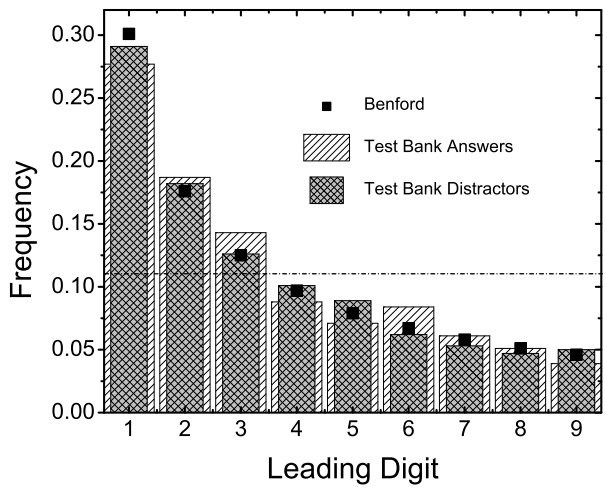}
\caption{The distribution of leading digits in multiple-choice testbank answers and associated distractors for Knight, ``Physics for Scientists and Engineers", 3rd edition. The dashed horizontal line indicates uniform distribution of first digits. Statistical analysis confirms conformation to Benford's Law, overlaid as black squares. Conformation of the distractors to Benford's Law precludes a Benford-based attack on the testbank.}
\label{fig:testbank}
\end{figure}

\begin{figure}
\includegraphics[width=\columnwidth]{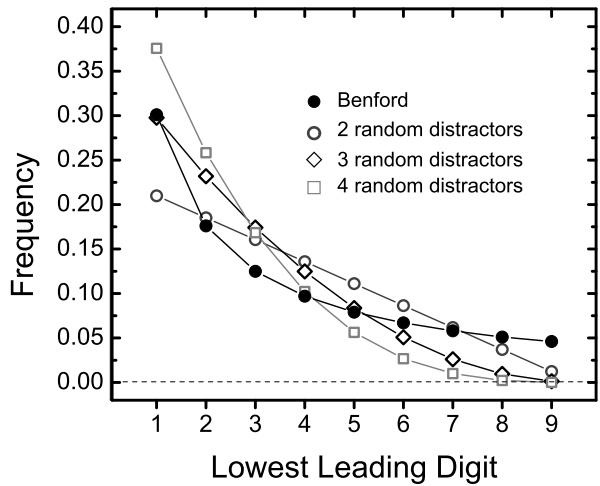}
\caption{Benford's Law and leading digit distribution from an ensemble of uniformly-distributed-first-digit distractors for 3-, 4-, and 5-option multiple choice questions. For a Benford's Law-based attack on a testbank the predominance of low-value leading digits in the answers must be maintained in the presence of a group of distracotrs. Despite the fact that for 4- and 5-option questions the distractors are collectively more likely to have the lowest leading digit, a Benford attack on such a group is nonetheless expected to yield an advantage over a random-guessing strategy. In the case of a test with 3-option questions---where the answers are Benford distributed and the two distractors are uniformly distributed---a Benford attack is expected to yield a passing score of 51$\%$.}
\label{fig:random_distractors}
\end{figure}

The first-digit frequency distribution for the companion multiple-choice testbank to Knight is presented in Fig. \ref{fig:testbank}. The keyed-responses (i.e. correct answers) conform acceptably to Benford's Law, yielding a MAD value of 0.011. Thus, over 50\% of the answers to numerical multiple-choice exam questions are anticipated to have a leading digit of 1, 2, or 3. As mentioned above, a Benford's Law-based attack on a multiple-choice testbank would be guided by the assumption that it is more likely that the incorrect options (distractors) are semi-randomly selected in such a way that each yields a uniform first-digit distribution. Then the question remains whether the latent Benford distribution of the keyed response will provide more low-digit options than an ensemble of uniformly distributed distractors. The computed expected distributions of \emph{lowest}-leading-digits in a group of distractors for each of a 3-, 4-, and 5- option multiple choice test is presented in Fig. \ref{fig:random_distractors}, overlaid with the Benford-distributed correct response.  

From Fig. \ref{fig:random_distractors}, we see that while for a 3-option test the lowest-leading-digits of 1 is more probable in the Benford-distributed keyed responses than in the combined pair of two distractors, at least one distractor in each 4- and 5-option item is expected to have \emph{on average} a lower first digit than the keyed response. These considerations would strongly moderate the advantage of a Benford's Law-based attack on a set of multiple-choice questions, but it would not entirely negate this strategy. To demonstrate this we generated 5,000 mock multiple-choice questions each comprising a keyed-response with a Benford's Law probability of leading digit, and 4 distractors with  uniformly-distributed leading digits. To simulate a 3-option MC test, we only considered the keyed response and the first two distractors. For a 4-option test we included the next distractor, and for a 5-option test we included the final distractor as well. We then identified the item with the lowest leading digit for each test type and scored a point if it was the keyed response. In the case of a tie among the keyed response and any number of distractors, we simply reverted to guessing among the ties. We find that such an attack always improves the test score over random guessing: For a 3-, 4-, or 5-option test we expect scores of 51\%, 41\%, or 33\%, respectively. Compared to blind-guessing scores of 33\%, 25\%, and 20\%, a Benford attack promises a significant advantage. The expectation of a passing score in a 3-option exam is particularly noteworthy considering expert recommendations that this is (psychometrically) the most desirable type of multiple-choice item \cite{rodriguez_2005}. 
\par We attacked the Knight testbank by selecting the option with the lowest leading digit or guessing among items with identical lowest leading digits, and compared these selections to the keyed response. The majority of questions were of the 4-option type. Our attack yielded a score of 24.6\%; no better than chance. The explanation for the failure in this strategy lies in the first-digit distribution of the distractors. As shown in Fig. \ref{fig:testbank} (and listed in Table \ref{tab:data}), the testbank distractors also closely conform to Benford's Law, passing the strictest MAD test. Thus, the fact that the keyed responses are Benford distributed is marginalized by the likewise distributed distractors.

\section{Discussion}

As we have shown, typical physics and chemistry questions, as a group, follow Benford's Law for leading digits, as do both the keyed responses and the distractors of a large introductory physics testbank. Random numbers, however, do not follow Benford's Law, but rather have uniformly-distributed leading digits. Thus, one definite conclusion we can draw from our analysis of the testbank is that for this set of multiple-choice questions the distractors are clearly not random numbers. A more interesting question is whether the fact that the distractors follow the same pattern of leading digits as do the keyed-responses---which emerge from well-defined and deterministic procedures---means that they too are a result of a similar creation process? That is, are the distractors necessarily created as answers to alternate questions? Not necessarily. While \emph{de novo} random numbers are uniformly distributed in leading-digit, processed random numbers are not, and multiplications of arrays of random number are known to generate near-ideal Benford sets \cite{adhikari_1968}. Furthermore, a random selection of numbers from multiple sets of different distributions---none of which needs be Benford distributed---yields a Benford set. This appears to be a leading-digits analogue to the central limit theorem \cite{benford_1938, hill_1995}. Finally, the scale invariance of Benford's Law suggests that multiplying values within a Benford set by various other numbers maintains the distribution of first digits. Thus, we can not identify the way in which the distractors were created. Whether they are made of processed random numbers, processed from the keyed response, or are themselves (perhaps erroneous) answers to a set of questions cannot be discerned from their distribution of leading digits.  
\par The question of why a large but seemingly random subset of all possible quantitative physics questions should follow Benford's Law with such precision is warranted. There may not be a clear-cut answer to this question, but the truth probably lies in the aforementioned theorem that random sampling from a wide mixture of first-digit distribution sets converges to a Benford distribution. As a group, end-of-chapter questions (or potential final examination questions) span many topics and involve numerous different parameters, each of which may have a different first-digit distribution within the domain of physically relevant phenomena. As examples, the sets of likely ``kinetic energies for vehicles on earth" and realistic ``currents induced in copper rings by Lenz's Law" may be sufficiently limited in domain as to individually refrain from a Benford distribution, but random sampling of values from such distributions will yield a Benford set. This is likely the reason that such a good Benford distribution is found in our samples. As further evidence of this argument, we see that combining the data from the three textbooks, the testbank answers, and the testbank distractors yields an aggregate data set that better conforms  to Benford's Law than any of the individual data sets. Included in Table \ref{tab:data}, this combined dataset shows a minuscule MAD of 0.0033. \\

Thus far we have determined that answers to physics questions conform to Benford's Law. However, these numerical quantities, as found in the textbooks and in the testbank, are often reported to an artificially reduced number of significant digits. The link between rounding and Benford's Law was identified at inception,\cite{benford_1938} and has formed the basis for some potential applications of the law for computer design \cite{pinkham_1961, barlow_1985}. Typical datasets reported in the Benford's Law literature contain many more than three significant digits. For pedagogical reasons, numbers reported in introductory physics courses are limited to the fewest number of significant digits warranted by the precision of the values used for the calculation or measurement. Instructional material with numbers that are reported to two or three significant figures may thus represent an artificial data set with an imperfect Bendord distribution of leading digits.  A description of how rounding modifies the distribution of leading digits is fairly straightforward: In the case of first-digit frequencies, the effect of rounding is to diminish the expected number of 1s and increase the frequency of all other digits. We present a modification to Eq.\ref{eq:benford} that includes the effects of rounding, where $N_{SD}$ is the number of significant digits to which a value is rounded, and $N_{SD}=\infty$ represents the traditional case of no rounding: 

\begin{equation} \label{eq:BLrounding}
\text{Prob}\left(D_i\:;\:N_{SD}\right) =
\begin{cases}
 \begin{split}
 \log\left(\frac{D_i+1-\frac{1}{2}10^{-\left(N_{SD}-1\right)}}{D_i-\frac{1}{2}10^{-\left(N_{SD}\right)}}\right) \\ D_i=1 \\  \\
 \log\left(\frac{D_i+1-\frac{1}{2}10^{-\left(N_{SD}-1\right)}}{D_i-\frac{1}{2}10^{-\left(N_{SD}-1\right)}}\right) \\ D_i\in \left\{2,3,...,9\right\}
\end{split}
\end{cases}
\end{equation}

\begin{table}
\caption{Comparing Benford distributions for leading digit in datasets with numbers rounded to one and two significant digits}
\bgroup
\def\arraystretch{1.1}
{\setlength{\tabcolsep}{0.5em}
\begin{tabular}{r || p{1.8 cm} c  c }
 \hline \hline
$D_{i}$ & \textbf{Benford's Law} ($\mathbf{N_{SD}=\infty}$) & $\mathbf{N_{SD}=2}$ & $\mathbf{N_{SD}=1}$ \\
 \hline \hline
 \textbf{1} & 0.301 & 0.292 & 0.198 \\
 \hline
 \textbf{2} & 0.176 & 0.180 & 0.222 \\
 \hline
 \textbf{3} & 0.125 & 0.127 & 0.146 \\
 \hline
 \textbf{4} & 0.0969 & 0.0980 & 0.109 \\
 \hline
 \textbf{5} & 0.0792 & 0.0799 & 0.0872 \\
 \hline
 \textbf{6} & 0.0669 & 0.0675 & 0.0726 \\
 \hline
 \textbf{7} & 0.0580 & 0.0584 & 0.0622 \\
 \hline
 \textbf{8} & 0.0511 & 0.0515 & 0.0544 \\
 \hline
 \textbf{9} & 0.0458 & 0.0460 & 0.0483 \\
 \hline \hline
 \multirow{2}{*}{\textbf{MAD}} & {} & 0.0020 & 0.023 \\
 &{} & close conform & non-conform \\
  \hline \hline
 \end{tabular}}
\egroup
\label{tab:rounding}
\end{table}

For numbers rounded to three or more significant digits the distribution of the leading digit is nearly identical to Benford's Law. However, in the case of values rounded to two significant digits, there is a small but significant modification of the first few digits. If numbers are rounded to one significant digit, the distribution of this leading digit varies drastically from Benford's Law to the point where the occurrence of the digit 2 becomes more probable than the digit 1. Table \ref{tab:rounding} summarizes this relationship, showing that a dataset comprised of values rounded to two significant digits would pass a test of conformity to Benford's law, with $\text{MAD}=0.0020$, but a dataset of values rounded to a single significant digit would fail to conform to a Benford's Law test despite the conformity of the underlying phenomena (i.e. the un-rounded data). We have observed that approximately 30\% of entries in the testbank are reported to two significant digits. Thus, perhaps for the testbank a more appropriate distribution of leading digits than given by Eq.\ref{eq:benford} should be given by a respective 0.7:0.3 weighted average of Eqns.\ref{eq:benford} and \ref{eq:BLrounding}. When compared to this hybrid distribution, we obtain a MAD value of 0.0040, which is slightly better than the MAD value of 0.0045 of this full testbank dataset to the unmodified Benford's Law. Thus, we are likely observing the effects of rounding in our data, and this effect is expected to be a factor in the leading-digit analysis of any similar dataset where rounding to one or two significant figures is common practice. 

\section{Summary and Conclusions}

We recorded the leftmost significant digit of the answers to every end-of-chapter question in two popular introductory physics textbooks, an intermediate analytical chemistry textbook, and a large introductory physics multiple-choice testbank, and find that all conform to Benford's Law. The fact that the answers to multiple-choice testbank questions follow this trend suggested a means by which the testbank could be attacked by a subject-ignorant but test-wise student. We find that among a set of distractors (wrong answers), each having uniformly distributed leading digits, the Benford distribution of the keyed response could be used to pass a 3-option test. Nonetheless, when using this information to ``guess" the correct answers in a real testbank we find that the distractors themselves conform to Benford's Law, thereby securing the testbank from such an attack. We observed that physics textbooks and testbank items are often reported rounded to two significant digits, and we have shown how this fact may have impacted its distribution of leading digits, modifying it slightly from Benford's law and primarily yielding a relative dearth of leading 1s. We expect that for many readers our demonstration of Benford's law in end-of-chapter textbook answers will prove counter-intuitive, as most people tend to believe that the leading digits of random values would be uniformly distributed. Armed with the knowledge that answers to physics questions follow Benford's law, one trifling piece of advice we can give to the test-wise student is as follows: At the end of a long constructed-response examination, if you have little time to double-check the answers to all of the questions, spend time on those questions that yielded final answers that have the largest leading digits; questions are expected to have answers with leading digits 7, 8, or 9 only 15\% of the time.

\begin{acknowledgments}
We thank Emily Slepkov for useful discussion and manuscript editing. We thank Pearson Education for the introductory physics multiple-choice testbank.  
\end{acknowledgments}

\bibliography{Benford_abrev}{}

\begin{thebibliography}{}
\bibitem{chi2_note}
Indeed, in our study we find the $\chi^2$ test to suffer from such excess
  power. Generally we find that our small data sets such as Knight and Skoog
  end-of-chapter answers pass the $\chi^2$ threshold test (15.5 for p\textless
  0.05 and 8 degrees of freedom), while the larger data sets such as the
  testbank distractors and Young end-of-chapter answers marginally fail the
  test despite the fact that they visually appear to conform to Benford's Law
  as well or better than the former.
\end{thebibliography}
\bibliographystyle{ajp}

\end{document}